\documentclass[
  twocolumn,
  prb,
  showpacs,
  amsmath,
  amssymb,
  superscriptaddress,
  floatfix
]{revtex4}

\usepackage{bm}
\usepackage{graphicx}
\usepackage{color}

\usepackage{ulem}

\newcommand{\be}{\begin{equation}}
\newcommand{\ee}{\end{equation}}
\newcommand{\beml}{\begin{subequations}}
\newcommand{\eml}{\end{subequations}}
\newcommand{\bea}{\begin{eqnarray}}
\newcommand{\eea}{\end{eqnarray}}
\newcommand{\ba}{\begin{array}}
\newcommand{\ea}{\end{array}}
\newcommand{\bpm}{\begin{pmatrix}}
\newcommand{\epm}{\end{pmatrix}}

\DeclareMathOperator{\im}{Im}
\DeclareMathOperator{\re}{Re}

\DeclareMathOperator{\lit}{li_4}

\begin{document}

\title{\hspace*{-0.5cm} Tunneling into the localized phase near Anderson
transitions
with Coulomb interaction}

\author{I.S.~Burmistrov}

\affiliation{ L.D. Landau Institute for Theoretical Physics, Kosygina
  street 2, 119334 Moscow, Russia}
\affiliation{Moscow
Institute of Physics and Technology, 141700 Moscow, Russia}

\author{I.V.~Gornyi}
\affiliation{
 Institut f\"ur Nanotechnologie, Karlsruhe Institute of Technology,
 76021 Karlsruhe, Germany
}
\affiliation{
 A.F.~Ioffe Physico-Technical Institute,
 194021 St.~Petersburg, Russia.
}

\author{A.D.~Mirlin}
\affiliation{
 Institut f\"ur Nanotechnologie, Karlsruhe Institute of Technology,
 76021 Karlsruhe, Germany
 }
\affiliation{
 Institut f\"ur Theorie der kondensierten Materie and DFG Center for Functional Nanostructures,
 Karlsruhe Institute of Technology, 76128 Karlsruhe, Germany
}
\affiliation{
 Petersburg Nuclear Physics Institute,
 188300 St.~Petersburg, Russia.
}

\begin{abstract}
We study the tunneling density of states (TDOS) of a disordered
electronic system with Coulomb interaction on the insulating side of the
Anderson localization transition. The average TDOS shows a
critical behavior at
high energies, with a crossover to a soft Coulomb gap $\Delta$ at low energies. We demonstrate that the
single-particle excitations experience a localization transition (which belongs to the noninteracting universality
class) at an energy $E=\pm E_c$. The mobility edge $E_c$
scales with the distance $\mu_c-\mu$ from the interacting critical point
according to $E_c\propto (\mu_c-\mu)^{\nu z}$, where $\nu$ and $z$ are the
localization-length
and the dynamical critical exponents. Local
TDOS shows strong fluctuations and long-range correlations which reflect
the multifractality associated with interacting and noninteracting fixed
points as well the localization of low-energy excitations.
\end{abstract}


\maketitle

\section{Introduction}%

Anderson localization
\cite{AL50} is one of major research fields in the condensed
matter physics. Of particular interest are Anderson
metal-insulator transitions (MITs) which are quantum phase transition with
a variety of remarkable properties. The physics of such transitions is by now
well
understood \cite{Evers08} in situations when the electron-electron
interaction is irrelevant in the renormalization group (RG) sense. This is the
case in problems with a short-range interaction and broken spin-rotation symmetry. The
case of a long-range ($1/r$) Coulomb interaction is more complicated.
A perturbative analysis based on a diffuson diagram technique \cite{AA}
showed that the interaction affects the conductivity in an essential way and
may thus be important for the physics of a MIT.
Initial scaling ideas \cite{mcmillan81,gefen83} emphasized the mutual influence of
interaction and disorder. A systematic theoretical approach to the problem
was built in the framework of the
non-linear $\sigma$-model (NL$\sigma$M) formalism. \cite{Finkelstein1983} This has
allowed to develop a scaling theory of the transition, \cite{finkelstein90,belitz94}
in agreement with experimental observations of
a MIT in three-dimensional (3D) semiconductor structures. \cite{semicond}

The goal of the present paper is  to explore the nature of single-particle excitations 
in a Coulomb-interacting disordered system that is slightly on the localized side of the Anderson MIT.  
The character of excitations is revealed by correlations and fluctuations of the local tunneling density of states (TDOS). 
In particularly, we study a fate of the single-particle mobility edge in the presence of Coulomb interaction and 
manifestation of localization of excitations in TDOS.

Experimental studies of the average TDOS \cite{massey96,lee99,teizer00,lee04} have
demonstrated that at the transition point it vanishes in a
power-law fashion, $\langle\rho(E)\rangle\sim |E|^\beta$ (energy $E$ is counted from the chemical potential $\mu$), in agreement with the
theoretical prediction. \cite{mcmillan81,Finkelstein1983,gefen83} Further, it was found
that on the localized side of the transition an additional soft gap with a
stronger suppression of the average TDOS opens around the Fermi energy.
To describe the TDOS near the MIT, a scaling ansatz was put forward in Ref. [\onlinecite{lee99}]
(see also [\onlinecite{amini13}]). With assumption that the exponent $\beta$ is determined by the dynamical exponent $z$, 
the scaling ansatz extrapolates~\cite{lee99} the physics of a soft Coulomb gap from the insulating regime \cite{ES}
to the criticality, where it agrees with the results of Ref.~[\onlinecite{mcmillan81}].
A similar behavior was found in a Hartree-Fock (HF) modeling of
the problem. \cite{amini13} However, while qualitatively capturing observed features of the TDOS, the approach
of Refs. [\onlinecite{mcmillan81,lee99}] does not fully reflect the complexity of the
problem since, in general, the exponents $\beta$ and $z$ are independent as seen for the Anderson MIT in the spatial dimensionality $d=2+\epsilon$. \cite{Finkelstein1983}

A further experimental motivation for our work is a recent
paper [\onlinecite{richardella10}] where TDOS near a MIT was studied locally by the
scanning tunneling microscopy (STM) approach. Strong fluctuations and
long-range correlations of the TDOS were found there and analyzed in terms
of multifractality, which was supported recently by numerical analysis in the framework of the density functional theory, \cite{Slevin} the HF simulation, \cite{amini13} and the NL$\sigma$M analysis. \cite{burmistrov13} While
Ref.~[\onlinecite{richardella10}] focused on the critical point and the metallic
side of the transition, it should be possible to extend an experimental study
of fluctuations and spatial correlations of local TDOS to the
insulating side.

\section{Diffusion and localization near the mobility edge}

According to the RG analysis of the NL$\sigma$M
\cite{finkelstein90,belitz94},
scaling properties of the diffusion constant $D(\omega,q)$  at the critical
point are 
\begin{equation}
 \label{e1}
 D(\omega,q) \sim \left\{ \begin{array}{cc}
                q^{d-2}, \qquad & \omega \ll q^z, \\
                \omega^{(d-2)/z}, \qquad & \omega \gg q^z,
               \end{array}
\right.
\end{equation}
where $z$ is the dynamical exponent. We assume below that $z<d$, which is normally the
case. 

Equation (\ref{e1}) discards a possible effect of
multifractality on the diffusion constant $D(\omega,q)$ in the large-momentum
range, $q^z \gg \omega$. In fact, it is known that in the noninteracting
problem, the multifractality does influence $D(\omega,q)$ of a critical system
at $q^d \gg \omega$. \cite{Chalker88,Evers08} Further, it has been recently shown \cite{burmistrov13}
that multifractality of local TDOS exists also in the presence of Coulomb
interaction (see also Ref. [\onlinecite{amini13}], where the multifractality 
was found within the HF approximation). It is thus plausible that the
multifractality affects the $q\gg \omega^{1/z}, \xi_0^{-1}$ behavior of the
diffusion constant in the interacting case as well (see Appendix \ref{Sec1}).
It can be shown, however, that multifractality would
have no influence on our results on the scaling of
average TDOS, since none of them are controlled by this range of momenta
and frequencies (see Appendix \ref{AppTDOS}).

Since the compressibility $\partial n/ \partial
\mu$ entering the Einstein relation is
not critical, Eq.~(\ref{e1}) determines also the scaling of conductivity at
criticality. 
While our main interest is in the 3D case, it is useful to consider a general $d>2$, since critical exponents can be controllably calculated in $d=2+\epsilon$ dimensions.

We will consider the localized phase, $\mu < \mu_c$
(here $\mu$ is the chemical potential),
near the transition, where the localization length
\begin{equation}
\label{e1aa}
\xi_0 \sim (\mu_c - \mu )^{-\nu}
\end{equation}
is finite but large.
In this case Eq.~(\ref{e1}) is applicable under the conditions
$\omega\gg\xi_0^{-z}$. In the opposite regime, $\omega\ll\xi_0^{-z}$ the system gets
localized. Thus, $D(\omega,q)$ has a behavior characteristic of the localized
regime
\begin{equation}
 \label{e1a}
 D(\omega,q) \simeq -i\omega P (q),
\end{equation}
which means that the propagating density has a finite long-time limiting
shape. Matching Eq.~(\ref{e1}) with (\ref{e1a}) at $\omega\sim\xi_0^{-z}$, we
find
\begin{equation}
 \label{e1b}
 P (q) \sim
  \left\{ \begin{array}{cc}
                q^{d-2} \xi_0^z, \qquad & q \gg \xi_0^{-1}, \\
                \xi_0^{z+2-d}, \qquad & q \ll \xi_0^{-1}.
               \end{array}
\right.
\end{equation}
The electric susceptibility $\chi(\omega,q)$ (determining the
permittivity via $\varepsilon=1+4\pi\chi$) is given in 3D by
\begin{equation}
\label{e2}
\chi(\omega,q) = {1\over 4\pi}V_0(q) \Pi(\omega,q) = {e^2 \over q^2}
{\partial
n\over \partial\mu} {D(\omega,q)q^2 \over D(\omega,q)q^2 - i\omega},
\end{equation}
where $V_0(q) = 4\pi e^2/q^2$ is the bare Coulomb
interaction and $\Pi(\omega,q)$ the polarization operator.
In the frequency range corresponding to the
localized regime, $\omega\ll\xi_0^{-z}$, we substitute (\ref{e1a}) into
(\ref{e2}) and obtain
\begin{equation}
\label{e2a}
 \chi(\omega,q) = e^2 (\partial
n /\partial\mu) P (q)/[ 1 + P (q)q^2].
\end{equation}
In the low-momentum range, $q\lesssim\xi_0^{-1}$, the second term in the
denominator can be neglected. Using Eq.~(\ref{e1b}), we find
the static, long-scale polarizability in 3D:
\begin{equation}
\chi \sim \xi_0^{z-1},
 \label{e2c}
\end{equation}
in agreement with Ref.~[\onlinecite{finkelstein90}].
A similar analysis for arbitrary  $d$ (with
$V_0(q)\sim q^{1-d}$) yields $\chi\sim\xi_0^{z+2-d}q^{3-d}$ (see Appendix \ref{Sec1}).

\section{Disorder-averaged TDOS}%
 
We remind that suppression of the disorder-averaged TDOS in a diffusive interacting system has
a form of a generalized Debye-Waller factor,
\begin{equation}
 \label{e3}
\langle \rho(E) \rangle = \rho_0 T \, {\rm Im} \int_{-{1/T}}^{1/T}
d\tau \, \frac{e^{i\epsilon_n \tau-J(\tau)}}{\sin \pi T\tau} \, \Big
|_{i\epsilon_n \to E+i0}\,,
\end{equation}
reflecting the charge spreading affected by
gauge-type phase fluctuations. \cite{finkelstein90,nazarov89a,levitov97,kamenev99,mishchenko01,mora07,gutman09}
Here $\rho_0$ is a non-critical high-energy TDOS (differing from $\partial n
/\partial \mu$ only by Fermi-liquid corrections), $T$ stands for the temperature,  and
\begin{equation}
 \label{e4}
 J(\tau) =\frac{1}{\rho_0} \int {d^d q \over (2\pi)^d} T \sum_{\omega_m} \frac{(1-\cos \omega_m\tau) Z}{Dq^2 (Dq^2 + Z|\omega_m|)}
\, .
\end{equation}
Here the frequency renormalization factor \cite{Finkelstein1983} $Z(i\omega_m,q)$ has the following asymptotic behavior at $|\omega|\ll \xi_0^{-z}$: 
\begin{equation}
\label{e1aaa}
Z(\omega,q)  \sim 
 \left\{ \begin{array}{cc}
                q^{d-z}, \, & q\xi_0 \gg 1, \\
                \xi_0^{z-d}, \, & q\xi_0 \ll 1.
               \end{array}
\right.
\end{equation}
We use the imaginary-time formalism with
fermionic ($\epsilon_n$) and bosonic ($\omega_m$) Matsubara frequencies. Below we
focus on the zero-temperature limit.

At the MIT point Eq.~(\ref{e3}) leads to a power-law scaling,
\begin{equation}
\langle\rho(E)\rangle \propto |E|^\beta,
\label{e7}
\end{equation}
where $\beta =O(1)$ in $d=2+\epsilon$ dimensions. Specifically,
up to corrections of order $\epsilon$, one finds $\beta \simeq 1/2$,
$1/[4(1-\ln 2)]$, and 1, for problems with magnetic impurities, magnetic field,
and spin-orbit scattering, respectively. \cite{finkelstein90,belitz94}

When the system is slightly off criticality, Eq.~(\ref{e7}) is valid for
energies $|E| \gg \xi_0^{-z}$ where $\xi_0\sim|\mu-\mu_c|^{-\nu}$ is the
localization (for $\mu<\mu_c$) or correlation (for $\mu>\mu_c$) length. On the metallic side of the MIT, the
diffusion
coefficient $D$ approaches a finite limiting value at $|\omega|\ll \xi_0^{-z}$ and $q\ll\xi_0^{-1}$, so that the TDOS behavior (\ref{e7}) saturates at a constant
$\langle\rho(0)\rangle \rangle \sim \xi_0^{-z\beta}$. On the
insulating side, $D$ gets suppressed at $|\omega|\ll\xi_0^{-z}$
according to Eq.~(\ref{e1a}).  Thus, for Matsubara frequencies
$|\omega_m|\ll\xi_0^{-z}$ we get
\begin{equation}
 \label{e50}
 \int  \frac{d^d q}{(2\pi)^d} \frac{Z}{Dq^2 (Dq^2 + Z|\omega_m|)} \simeq \frac{2\Delta}
{\omega_m^2},
\end{equation}
where
\begin{equation}
 \label{e50rt}
 \Delta = \frac{1}{2} \int \frac{d^d q}{(2\pi)^d} \frac{Z}{P(q) q^2 (P(q) q^2 + Z)}  \sim \xi_0^{-z}.
\end{equation}
This yields a contribution $\Delta |\tau| $ to $J(\tau)$. The effect of such a
linear-in-time contribution is well known from the theory of Coulomb blockade:
\cite{nazarov89,devoret90,kamenev96} it opens a gap around
the Fermi energy, so that the LDOS $\langle\rho(E)\rangle$ vanishes for $|E|\leqslant \Delta$. 

We thus see that the system generates a gap in analogy with
the Coulomb blockade effect, with $\Delta$ playing a role of the effective
charging energy. Physically, this can be understood as follows. One can think
of the system as breaking in ``quantum dots'' of the size of the
localization length $\xi_0$.
We will show below that low-lying excitations (with energies well below
$\Delta$) are localized on a scale $\xi_0$. Now consider two such states with
energies $E_1>0$ (unoccupied) and $E_2<0$ (occupied) which are located within
the same (or nearby) ``quantum dots'' so that the distance between their
centers is $\sim \xi_0$. The screened Coulomb interaction between this states is
$\sim \xi_0^{-z}$ according to Eq.~(\ref{e2c}). This implies that the
difference $E_1-E_2$ should be $\gtrsim \xi_0^{-z}$; otherwise we would
reduce the energy by shifting an electron from the state 2 to the state 1,
cf. Efros-Shklovskii (ES) argument. \cite{ES}
Hence, there is a Coulomb gap of the size $\sim \xi_0^{-z} \sim \Delta$.

In fact, contributions to $J(\tau)$ from $|\omega_m|\gg \Delta$ as well as corrections to the approximation \eqref{e3}  smear the gap. \cite{nazarov89,devoret90,kamenev96}
In $d=2+\epsilon$ dimensions Eq. \eqref{e3} yields the following results  for the average TDOS: (i) the power-law scaling \eqref{e7} at $|E| \gg \epsilon\,  e^{z \nu} \Delta$, (ii) almost constant dependence on energy, $\langle \rho(E)\rangle \sim |\ln \ln (|E|/\Delta)|^{-z\nu\beta}$, for $\Delta \ll |E| \ll \epsilon\, e^{z \nu} \Delta$, and (iii) linear dependence on energy, $\langle\rho(E)\rangle \sim |E|$, for $|E|\ll \Delta$ (see Appendix \ref{AppTDOS}).

We note that linear energy dependence of the average TDOS is also obtained within an approximation of a hard gap of a fixed width $2\Delta$ from the following argument.
Let us assume that the precise position of the gap
may fluctuate from point to point (corresponding to a random value of a
potential on each ``dot''), with a restriction that the gap includes
the zero energy. \cite{efros76,skinner12}  If we neglect interaction on scales larger than
$\xi_0$, the ``dots'' will become uncorrelated, so that the center of the gap will
be uniformly distributed in the interval $[-\Delta,\Delta]$. This would yield an
average TDOS vanishing linearly at $|E| < \Delta$:
\begin{equation}
\langle\rho(E)\rangle \sim \langle\rho(\Delta)\rangle |E| / \Delta\,,
\label{eqrhoL}
\end{equation}
where $\langle\rho(\Delta)\rangle\sim\Delta^\beta$
is the TDOS (\ref{e7}) on the lower boundary of the high-energy
(critical) regime.

Interaction between distant ``dots'' leads to an additional suppression of
the average TDOS due to ES mechanism. \cite{ES} We focus in
this discussion on the 3D situation.
The interaction between such states at a distance $r > \xi_0$ is
$ U(r) \sim 1 / \chi r \sim 1 / \xi^{z-1} r \sim \Delta \xi / r$,
where we used Eq.~(\ref{e2c}).
An empty state with energy $E_i > 0$ and a filled state with energy $E_j < 0$
corresponding to the ``dots'' $i$ and $j$ should  satisfy the ``single-particle stability criterion'':
$ E_i - E_j > U({{\bf r}_i - {\bf r}_j})$.
This yields the ES law for the low-energy behavior of TDOS:
\begin{equation}
 \label{e7af}
 \langle\rho(E)\rangle \sim E^2 / \Delta^3 \xi_0^3.
\end{equation}
Note that Eq.~(\ref{e7af}) does not match the high-energy
behavior (\ref{e7}) at $|E|\sim\Delta$
in view of a difference between the exponents $z$ and $z_\beta=d/(\beta +1)$.
Thus, within the single-particle Coulomb gap picture,
there should be an intermediate regime between
the critical regime $E\gtrsim\Delta$ and the low-energy
behavior (\ref{e7af}). An alternative way to express this mismatch is to introduce an energy scale
\begin{equation}
 \label{e7ag}
 \delta = \frac{1}{\langle\rho(\Delta)\rangle \xi_0^3} \sim \xi_0^{\beta z - 3} \sim \Delta \xi_0^{3(z/z_\beta
- 1)}
\end{equation}
with a meaning of an excitation level spacing within the length $\xi_0$. The inequality $z \neq z_\beta = d/(\beta+1)$, which, as we expect, is a generic case, implies a parametric mismatch between $\delta$ and $\Delta$. Although we know that $z>z_\beta$ for MIT in $d=2+\epsilon$, \cite{finkelstein90,belitz94} it is not known which of the
exponents $z$ and $z_\beta$ is larger in $d=3$, so that we consider both possibilities.

(i) $z < z_\beta$, i.e. $\delta \ll \Delta$.\, This is a usual situation from the Coulomb-blockade point of view. The ``charging energy'' $\Delta$ is much larger than the level spacing $\delta$ in the
``dot'' (localized region of typical size $\xi_0$). The ES formula (\ref{e7af}) is valid only for $|E| < \delta$. In the
intermediate range $\delta < |E| < \Delta$ we should take into account
contribution of all excited states in the dot. This enhances the ES result by
a factor $|E|/\delta$, yielding
\begin{equation}
 \label{e7ah}
 \langle\rho (E)\rangle \sim \left\{
 \begin{array}{cc}
 E^2/(\Delta^3 \xi_0^3)\,, &\qquad |E| \ll \delta \, ,\\
 |E|^3/(\delta \Delta^3 \xi_0^3)\,, &\qquad \delta \ll |E| \ll \Delta \,.
 \end{array}
  \right.
\end{equation}
At $|E| \sim \Delta$ this matches the result \eqref{e7} for the disorder-averaged TDOS at the criticality.

(ii) $z > z_\beta$, i.e. $\delta \gg \Delta$.
This is a case opposite to the usual Coulomb-blockade situation: charging
energy is now much smaller than the level spacing (note that this situation is realized in $d=2+\epsilon$).
As a result, the ES
mechanism is not very efficient in suppressing the average TDOS, so that in the
most of the gap region it is expected to be given by Eq. \eqref{eqrhoL}.
Only at the lowest energies will the ES mechanism be operative, further suppressing the
TDOS. We thus expect that the TDOS in the gap will be given by the minimum of
(\ref{eqrhoL}) and (\ref{e7af}),
\begin{equation}
\label{e7ai}
 \langle\rho (E)\rangle \sim \left\{
 \begin{array}{cc}
E^2/(\Delta^3 \xi_0^3)\,, &\quad |E| \ll \Delta^2 /\delta \, ,\\
|E|/(\Delta \delta \xi_0^3)\,, &\quad \Delta^2 /\delta \ll |E| \ll \Delta \,.
 \end{array}
 \right.
\end{equation}

It is worth noting, however, that the single-particle stability criterion yields
only an upper bound for the TDOS. Many-body stability criteria
may lead to a stronger suppression \cite{efros76,ESred,efros11}
of the true TDOS in the limit $|E|\to 0$. 
Therefore, the results (\ref{e7ah}) and (\ref{e7ai}) for the average TDOS should be regarded
as the estimates from above.

\section{Localization transition for excitations}%

Now we explore the character of excitations and, in particular, the
dependence of the localization length $\xi$ on an excitation energy $E$. 
We remind that then $\xi_0$ is the zero-energy localization length, $\xi(E=0)
\equiv\xi_0$.

It is instructive to consider first the noninteracting case when the
mobility edge for excitations is simply $E_c=\mu_c-\mu$.
The critical behavior at this edge, when $E$
crosses $E_c$, is clearly the same as the zero-energy critical behavior with
$\mu$ driven through the transition point $\mu_c$. As we discuss
now, the case of a Coulomb-interacting system differs from this picture in
several crucial aspects.

\begin{figure}[t]
\centerline{(a) \includegraphics[width=0.8\columnwidth]{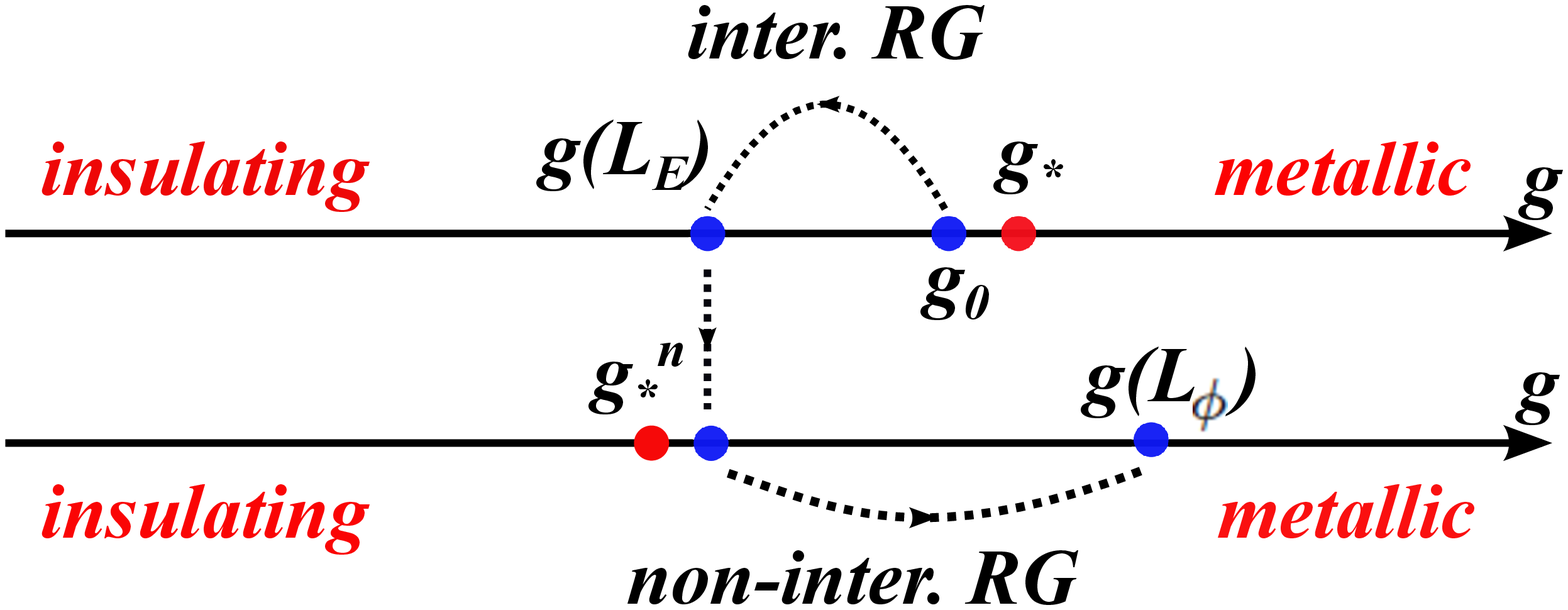}}
\vspace{0.5cm}
\centerline{(b) \includegraphics[width=0.8\columnwidth]{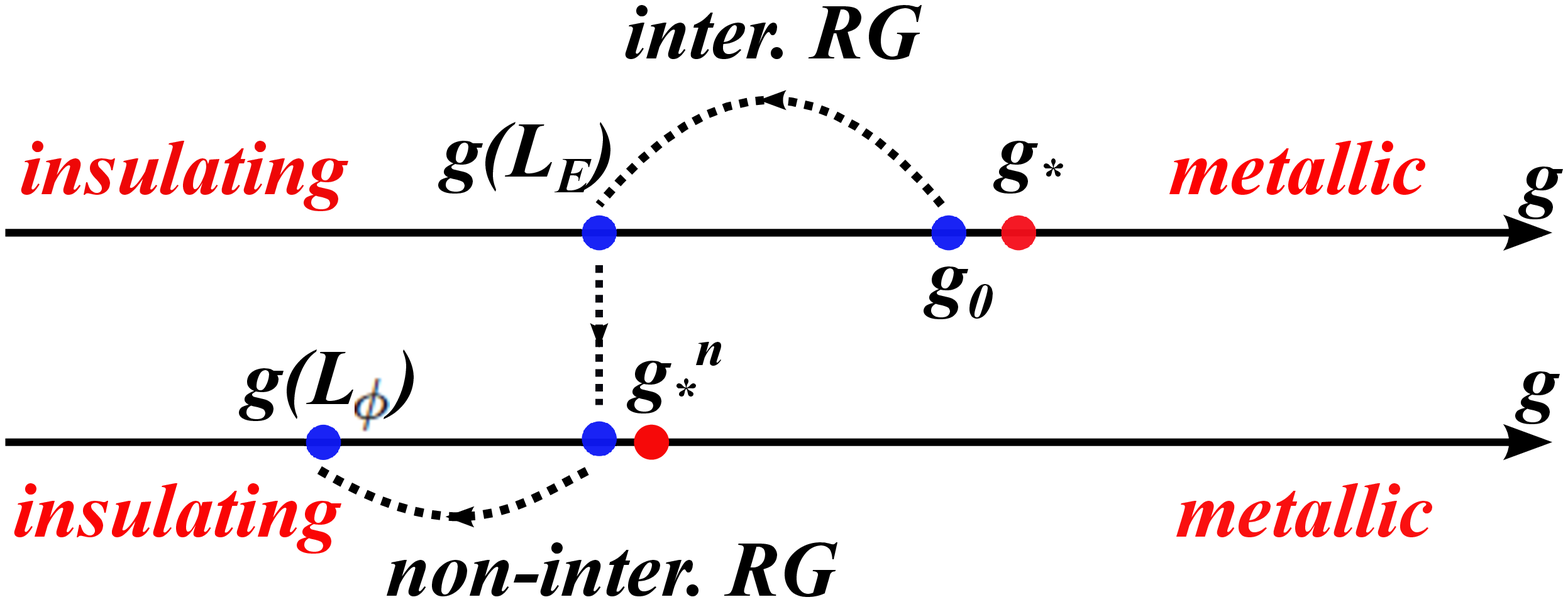}
}
\caption{(Color online) Sketch of the two-step RG in the case $g_*>g_*^{\rm n}$ for (a)  $|E|\gg \xi_0^{-z}$ and (b)  $|E|\ll\xi_0^{-z}$; $g_0$ denotes the bare value of $g$.}
\label{fig2}
\end{figure}

To identify the localization threshold $E_c$ for the interacting problem, we
proceed as follows. The
interacting NL$\sigma$M theory gets renormalized with a scaling factor $b$ according to
$q\to bq$, $\tau\to b^{1/\nu}\tau$,
and $E\to b^z E$,
where $\tau=(g-g_*)/g_*\sim(\mu-\mu_c)/\mu_c$ is a deviation of the
conductance from the interacting fixed-point value $g_*$. If
$|E|\gg\xi_0^{-z}$, the RG for the operators characterizing the
physics at an energy $E$ (e.g., moments of the LDOS)
proceeds in two steps. The first step is described by the interacting RG and
stops at the length scale $L_E\sim|E|^{-1/z}\ll\xi_0$. The output value of
$\tau$ is small (i.e. $g(L_E)$ is close to $g_*$): 
\begin{equation}
\tau = (\xi_0^{-z}/|E|)^{1/\nu z} \ll 1 .
\end{equation}
The second step of RG (between $L_E$ and $\min\{\xi,L_\phi\}$ where $L_\phi$ is the dephasing length) develops in accordance with the noninteracting theory
with $g(L_E)$ serving as the input value. We assume that the noninteracting
fixed point is characterized by a critical value $g_*^{\rm n} < g_*$. This is
certainly the case for problems in which exchange interaction is suppressed such that
the (direct) interaction enhances
localization. Then for energies well above $\xi_0^{-z}$ the conductance $g(L_E)$ is
above $g_*^{\rm n}$, so that the second step of RG starts at the metallic side of the noninteracting fixed point (see Fig. \ref{fig2}a). Therefore, the states with $|E|\gg\xi_0^{-z}$
are delocalized. With decreasing $|E|$ (increasing $L_E$), the value of $g(L_E)$ decreases such that at $|E|\ll \xi_0^{-z}$ the conductance $g(L_E)$ becomes below $g_*^{\rm n}$, and the second step of RG starts at the insulating side
of the noninteracting critical point (see Fig. \ref{fig2}b). The energy $E_c$ at which $g(L_E)$ reaches the value $g_*^{\rm n}$ scales as 
\begin{equation}
\label{e11d}
E_c =  \xi_0^{-z} \left (\frac{g_*-g_*^{\rm n}}{g_*} \right )^{\nu z}  \sim (\mu_c -\mu)^{\nu z} \,.
\end{equation}

When deriving the result (\ref{e11d}),
we have neglected the dependence of the bare diffusion constant (the bare $g$) on energy. This
is justified provided $ \nu z > 1 $,
which is satisfied both
theoretically (in $d=2+\epsilon$) and experimentally (in $d=3$). Taking into account the energy dependence of the bare $g$ would thus only slightly change the initial value $\tau_0=(g_0-g_*)/g_*$, without any essential effect.

\begin{figure}[t]

\centerline{\includegraphics[width=\columnwidth]{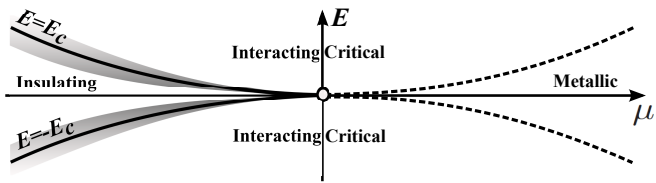}}

\caption{``Phase diagram'' in the $\mu$---$E$ plane.
The interacting MIT critical point is at $\mu=\mu_c$, $E=0$; it determines the
quantum critical behavior with the exponents $\nu$, $z$, $\beta$, $\Delta_q$.
The thick lines $|E|=E_c\sim(\mu_c-\mu)^{\nu z}$ emanating from this point
on the insulating
side, $\mu < \mu_c$, are lines of the Anderson transition for excitations.
This transition is characterized by the noninteracting critical exponents
$\nu_{\rm n}$, $\Delta_q^{\rm n}$; the corresponding critical regions are
shaded.
The average TDOS shows a critical scaling (\ref{e7}) above $E_c$ and a soft
Coulomb gap below $E_c$. A counterpart of $E_c$ on the metallic side
(dashed lines) marks a crossover from the critical to metallic
behavior. }

\label{fig1}

\end{figure}

The energy $E_c$ is the mobility edge for single-particle excitations. Let us emphasize that, contrary to the noninteracting problem, there is not only a mobility edge at $E_c$ for particles but also a second mobility edge at
$-E_c$ for holes. For $|E|<E_c$, the excitations are localized. In this region the
zero-temperature dephasing rate (or, equivalently, inelastic decay rate) vanishes. The excitations with $|E| >
E_c$ are delocalized. Under the condition $\nu z > 1$, the mobility edge $E_c$  satisfies $E_c\ll\mu_c-\mu$. 
The corresponding phase diagram is sketched in Fig.~\ref{fig1}. A qualitatively similar phase diagram was obtained in Ref. [\onlinecite{amini13}]
from the HF numerical modeling of the interacting Anderson transition on a 3D cubic
lattice of size $10^3$.

In $d=3$ the critical conductances  $g_*$ and $g_*^{\rm n}$ are of the order unity. Therefore, 
the width $\Delta$ of the soft Coulomb gap defined by Eq. \eqref{e50} is 
parametrically the same as the mobility edge $E_c$. Thus, one can roughly say
that the states inside (outside) the gap are localized (respectively,
delocalized). For MIT in $d=2+\epsilon$ in the presence of magnetic impurities the following results are known: $g_*^{\rm n} = 1/(\pi \sqrt{2\epsilon})$, $g_* =2/(\pi \epsilon)$, $\nu =1/\epsilon$ and $z= 2+\epsilon/2$.  
\cite{finkelstein90,belitz94} Hence, from Eqs. \eqref{e50} and \eqref{e11d} we find $E_c \sim \xi_0^{-z} \exp(1/\sqrt{2\epsilon})$ and $\Delta \sim \xi_0^{-z}/\epsilon$. Thus, while $E_c$ and $\Delta$ have the same scaling with the distance from the critical point, $E_c \sim \Delta \sim (\mu_c-\mu)^{\nu z}$, their dependence on $\epsilon$ is parametrically different. As a result, for $\epsilon\ll 1$ we find $E_c\gg\Delta$ implying that localized excitations exist far beyond the soft Coulomb gap $\Delta$.

A crucial consequence of the above analysis is that the localization transition
for excitations at $|E|=E_c$ is in the {\it noninteracting} universality class.
In particular, the localization length $\xi(E)$ scales near $E_c$ as
\begin{equation}
 \label{e11e}
 \xi(E) \sim \xi_0 \bigl (\bigl | E_c-|E| \bigr |/E_c\bigr )^{-\nu_{\rm n}}\,, \ \ \bigl | E_c-|E|\bigr | \ll E_c \,,
\end{equation}
where $\nu_{\rm n}$ is the exponent of the noninteracting
theory.

An important characteristic of excitations is their dephasing
length $L_\phi$. In the localized regime, $|E|<E_c$, there is no inelastic decay,
i.e., $L_\phi=\infty$. When energy $|E|$
approaches $E_c$ from above,  $|E| > E_c$, the dephasing length diverges:
\begin{equation}
\label{e13}
L_\phi \sim L_E  \Bigl (\bigl (|E|-E_c\bigr )/E_c\Bigr )^{-1/z_\phi^{\rm n}}, \ \ 
|E|-E_c \ll E_c.
\end{equation}
This is because the decay is only possible in the continuous spectrum: a
particle with energy $E>E_c$ can decay
into another delocalized particle with an energy $E'$ satisfying $E_c<E'<E$ and
a localized electron-hole pair. The
corresponding phase volume tends to zero when the energy approaches $E_c$ from
above. A Fermi golden-rule type calculation yields $z_\phi^{\rm n} = \max\{d^2/(4-d),d^2/(d+\Delta_2^{\rm n})\}$ (see Appendix \ref{AppDep}). Here $\Delta_q^{\rm n}$ stands for the noninteracting multifractal exponents. Since for the 3D Anderson transition in orthogonal symmetry class it is known from numerics that
$\Delta_2^{\rm n} = -1.7\pm 0.05$, \cite{Mildenberger2002} we obtain 
$z_\phi^{\rm n}=9$ in $d=3$.  For energies $|E|\gg E_c$ the system is at criticality from the point of
view of the interacting theory, so that the dephasing length is controlled by
the corresponding dynamical exponent, $ L_\phi \sim L_E \sim |E|^{-1/z}$.

\section{Fluctuations of TDOS}%
 
Fluctuations and correlations of TDOS in different energy ranges are
governed by three factors: (i) multifractality at the interacting fixed point,
(ii) multifractality at the noninteracting fixed point, and (iii) localization
of excitations below $E_c$.

For energies well above the mobility gap, $|E|-E_c\gtrsim E_c$, the
system is controlled by the interacting quantum critical point. Thus, the
moments of local TDOS show strong fluctuations governed by the corresponding
multifractal exponents $\Delta_q$: \cite{burmistrov13}
\begin{equation}
 \label{e15a}
 \langle \rho^q(E) \rangle / \langle \rho(E) \rangle^q \sim
L_{\phi}^{-\Delta_q}\,.
\end{equation}
These fluctuations are further enhanced for energies near the
mobility edge, $|E|>E_c$ and $|E|-E_c\ll
E_c$ such that the dephasing length $L_\phi$
is strongly enhanced and satisfies $L_\phi\gg\xi_0$. We note that $L_\phi$ in this regime is much shorter than the
correlation length $\xi(E)$ in view of $\nu_{\rm n}z_\phi^{\rm n}>1$. In this situation
the system shows interacting multifractal scaling up to the scale $\xi_0$ and
then noninteracting multifractality (with exponents $\Delta_q^{\rm n}$) up to
the scale $L_\phi$
\begin{equation}
 \label{e15b}
 \langle \rho^q(E) \rangle / \langle \rho(E) \rangle^q \sim
\xi_0^{-\Delta_q} (L_\phi/\xi_0)^{-\Delta_q^{\rm n}}\,.
\end{equation}

Since in the
localized regime, $|E|<E_c$, the dephasing length is infinite,
$L_\phi = \infty$, the LDOS
fluctuations will diverge,
$\langle\rho^q(E)\rangle=\infty$ for $q>1$. This result is regularized by a
small temperature  $T$ that yields a finite
(although large) dephasing length $L_{\phi T} \gg \xi_0$ due
to electron-electron and/or electron-phonon scattering processes. We
do not discuss different contributions here but simply consider $L_{\phi T}$ as
a parameter. For $T=0$ the regularization is provided by
the system size $L$.

When the energy is below the mobility edge but close to it, $|E| < E_c$ and $E_c
- |E| \ll E_c$, the system shows first the interacting
multifractal scaling up to
$\xi_0$, then the noninteracting multifractality up to $\xi$, and
finally insulator-like fluctuations up to the scale $L_{\phi T}$:
\begin{equation}
 \label{e15c}
 \langle \rho^q(E) \rangle / \langle \rho(E) \rangle^q \sim
\xi_0^{-\Delta_q} (\xi/\xi_0)^{-\Delta_q^{\rm n}} (L_{\phi T} /
\xi)^{d(q-1)}\, .
\end{equation}
When the energy is further lowered, $|E|<E_c$ and $E_c-|E|\sim E_c$, the
localization length $\xi$ becomes  of the order of $\xi_0$,
so that the second factor in Eq.~(\ref{e15c}) disappears.

The multifractality leads not only to strong fluctuations but also to
long-range spatial correlations of TDOS. In particular, at $|E|\gg E_c$ the correlation
function $\langle\rho(E,\bm{r})\rho(E,\bm{r}+\bm{R})\rangle$ shows a
power-law scaling $\sim R^{\Delta_2}$ up to the scale $\xi_0$. In the
vicinity of the mobility edge, $|E_c-|E||\ll E_c$ there is a range of distances,
$\xi_0\ll R\ll\xi$ (for $|E|<E_c$) or $\xi_0\ll R\ll L_\phi$ (for $|E|>E_c$)
where this correlation function shows the
scaling $\sim R^{\Delta_2^{\rm n}}$ controlled by the noninteracting
multifractal exponent.

\section{Discussion and conclusions}

We have studied the TDOS of a disordered
electronic system with Coulomb interaction on the insulating side of the
Anderson localization transition. The average TDOS shows a critical behavior at
high energies, with a crossover to a soft gap $\Delta$ at low energies. The latter regime
combines the physics of Coulomb blockade in quantum dots and that of Coulomb gap
deep in the insulating phase. The single-particle excitations experience a
localization transition at an energy $E=\pm E_c$. The mobility edge
$E_c$ and the soft Coulomb gap $\Delta$ show the same scaling with the distance from the critical point,
$E_c\sim\Delta\sim (\mu_c-\mu)^{\nu z}$, where  $\nu$ and $z$ are the critical
exponents of the interacting problem. The critical behavior of the
localization length of excitations near $E_c$ is controlled by
the exponent $\nu_{\rm n}$ of the noninteracting theory. Local
TDOS exhibits strong fluctuations and long-range correlations which reflect
the multifractality associated with interacting and noninteracting fixed
points as well with the localization of low-energy excitations.

It is worth discussing the relation of our findings to previous results on effects of interaction on localization properties of excitations. 

(i) In Refs. [\onlinecite{Imry1995,Shepelyansky1997}] the mobility edge for two-particle excitations above the Fermi sea ($E_{c2}$) 
was studied for the case of short range interaction. It was found that $E_{c2}$
is much lower than the naive mobility edge $\mu_c-\mu$. 
This implies that the true single-particle mobility edge is also lowered. Indeed, a single-particle excitation with energy
between $E_c$ and $E_{c2}$ can create an electron-hole pair and then use the excited electron to form a delocalized two-particle excitation.
We argue that in our problem with Coulomb interaction the mobility edge for two-particle excitations has the same scaling 
with the distance to the critical point as $E_c$ (see Eq. \eqref{e11d}). 
The reason for this is that in the case of Coulomb interaction all the energy scales relevant to diffusion 
behave as $\xi_0^{-z}$. In particular, the energy of Coulomb interaction for two closely located particles 
is given by $\Delta \sim E_c \sim \xi_0^{-z}$. Thus we expect that $E_c$ gives the correct scaling of the mobility edge for all types of charged excitations.

(ii) According to Refs. [\onlinecite{Anderson,FleishmanAnderson,Levitov1990}] Coulomb interaction leads unavoidably to delocalization 
of electron-hole pairs in dimensionality $d\geqslant d_c=3$ in view of slow decay of coupling between distant electron-hole pairs. 
This implies for our problem that localized charged excitations at $|E|<E_c$ may coexist with 
delocalized neutral excitations. 
Furthermore, the arguments based on consideration of processes involving four electron-hole pairs suggest that the critical dimension for delocalization of
electron-hole pairs is lower, $d_c=3/2$. \cite{Burin2006,Muller2013} 
We expect, however, that the presence of delocalized neutral excitations does not essentially affect our results for tunneling 
characteristics of the system since the latter are necessarily related to the transport of charge. A possible manifestation of the electron-hole delocalization would be 
a finite value of $L_\phi$ at zero $T$ in Eq. \eqref{e15c}. A more detailed study of the related
effects is relegated to future work.


The average TDOS and its fluctuations can be also experimentally studied in the
vicinity of 2D localization quantum phase transitions, including the
superconductor-insulator transition \cite{sacepe08} and the
quantum Hall transition. \cite{morgenstern-2D} Extending our analysis on these
transitions remains a challenging prospect for future research.

In a very recent špreprint [\onlinecite{Mottaghizadeh}], šMottaghizadeh {\it et al} explored the average TDOS in a memristive device with a tunable doping level across the MIT. Their findings are in an overall agreement with the ``phase diagram'' in our Fig. \ref{fig1} and with theoretical expectations for the corresponding regimes.

\begin{acknowledgements}

We are grateful to D. Gutman, 
V.E. Kravtsov, 
P. M. Ostrovsky, 
D.G. Polyakov, 
I.V. Protopopov, 
M.A. Skvortsov and B.I. Shklovskii for discussions.
The work was supported by DFG-RFBR in the framework of SPP 1285,
BMBF, Russian
President Grant No. MK-4337.2013.02, Dynasty Foundation, RFBR Grant No. 14-02-00333, by the Ministry of Education and Science of the Russian Federation, and by RAS Programs ``Quantum Mesoscopic and Disordered  Dystems'',
``Quantum Physics of Condensed Matter'', and ``Fundamentals of Nanotechnology
and Nanomaterials'', and by the Ministry of Education and Science of the Russian Federation.

\end{acknowledgements}


\appendix

\section{Screened interaction in the localized phase \label{Sec1}}

In the main text, we have focused on the screening properties of a 3D insulator. Here we present the results for the screened Coulomb interaction in an insulating phase near the metal-insulator transition (MIT) in 
an arbitrary spatial dimensionality $2<d\leqslant 3$. The screened electron-electron interaction $U(\omega,q)$ and the effective dielectric function $\varepsilon(\omega,q)$ are given as
\begin{equation}
U(\omega,q) = \frac{U_0(q)}{\varepsilon(\omega, q)}, \quad \varepsilon(\omega, q) =
1 + \frac{\partial n}{\partial \mu}  \frac{D q^2 U_0(q)  }{D q^2-i\omega} .
\label{eqUe}
\end{equation}
Here, ${\partial n}/{\partial \mu}$ is the thermodynamic density of states and $U_0(q) \sim q^{1-d}$ denotes the bare Coulomb interaction in $d$ spatial dimensions.
 It is worth mentioning that the irreducible  polarization operator which determines the screening in Eq. \eqref{eqUe} does not contain the Finkelstein's frequency renormalization parameter $Z(\omega,q)$ since it involves a standard (including the interaction vertex corrections) diffuson rather than a mesoscopic one. \cite{finkelstein90}

In the vicinity of the interacting critical point, $|g/g_*-1|\ll 1$ (we remind that $g$ denotes dimensionless conductance and $g_*$ stands for the critical value of the conductance), the diffusion coefficient can be written in the scaling form
\begin{equation}
D(\omega,q) = \xi_0^{2}(\xi_0/l)^{-d} \mathcal{R}_D(\omega/\Delta, q\xi_0), \quad
\Delta = E_0 (\xi_0/l)^{-z}.
\label{eqDcInt}
\end{equation}
Here $l$ and $E_0$ stand for the ultraviolet length and energy scales (the elastic mean-free path and the inverse elastic scattering time, respectively), the dynamical exponent $z$ relates the induced length scale $L$ with the energy $E$ or frequency $\omega$, $L \sim |E|^{-1/z}$ or $L \sim |\omega|^{-1/z}$, and $\xi_0 = l|g/g_*-1|^{-\nu}$ denotes the divergent localization/correlation length.

In the localized phase ($g<g_*$) the diffusion coefficient at low frequencies $\omega\ll \Delta$ can be written as 
\begin{equation}
D(\omega,q) =  -i\omega P(q) +  \left (e^2 \frac{\partial n}{\partial \mu}\right )^{-1} \re \sigma(\omega,q).
\label{eq10}
\end{equation}
To be consistent with the scaling form \eqref{eqDcInt} of the diffusion coefficient, the first (imaginary) term, which describes the polarizability of the system, should have the following asymptotic behavior:
\begin{equation}
P(q)\sim \xi_0^2 (\xi_0/l)^{z-d} \begin{cases}
(q \xi_0)^{d-2+\Delta_2}, &\quad q\gg \xi_0^{-1},\\
1, &\quad q\ll \xi_0^{-1} .
\end{cases}
\label{eq6}
\end{equation}
Here we include a possible effect of the multifractality in the presence of the Coulomb interaction \cite{burmistrov13} (see also Ref. [\onlinecite{amini13}]) on the diffusion coefficient which is characterized by the interacting multifractal exponent $\Delta_2<0$ 
(in the absence of this effect, $\Delta_2=0$).

In a noninteracting Anderson insulator, the real part of \textit{ac} conductivity
is given by Mott's formula, $\re \sigma(\omega) \propto \omega^2 \ln^{d+1}(\Delta/\omega)$.
However, in a Coulomb-glass insulator which we are dealing with, at frequencies $\omega\ll \Delta$
(as we shall demonstrate below, $\Delta$ is understood as the Coulomb potential at scale $\xi_0$) the Mott formula is modified
by Coulomb energy associated with the pairs of states involved in the transport. \cite{ES, ES1981, ESred}
The main modification is the replacement $\omega^2 \to |\omega| \Delta$ in the Mott formula. 
In what follows, we shall use the following expression for the real part of the \textit{ac} conductivity
\begin{equation}
\re \sigma(\omega) \sim e^2   \frac{\partial n}{\partial \mu} |\omega|^\alpha \Delta^{1-\alpha}   M(q) .
\label{eq11a}
\end{equation}
Here and in what follows, we disregard the logarithmic factors in $\re \sigma(\omega,q),$ since we will focus on the power-law dependencies. We use $\alpha=2$ for the Mott formula and $\alpha=1$ when the modifications due to the Coulomb interaction are taken into account according to Ref. [\onlinecite{ES, ES1981, ESred}].  The function $M(q)$ behaves as
\begin{equation}
M(q)\sim \xi_0^2 (\xi_0/l)^{z-d} \begin{cases}
(q \xi_0)^{d-2+\Delta_2}, &\quad q\gg \xi_0^{-1},\\
1, &\quad q\ll \xi_0^{-1} .
\end{cases}
\label{eq6a}
\end{equation}

It is convenient to introduce the inverse static screening length $\varkappa$ as
$({\partial n}/{\partial \mu})U_0(q) = (\varkappa/q)^{d-1}$. Hereinafter, we assume that the inequality
\begin{equation}
q\ll 1/l \ll \varkappa
\label{eqIneq}
\end{equation}
holds. Using Eqs.~(\ref{eq6}) and (\ref{eq6a}), we find the statically screened Coulomb potential in different domains of momenta:
\begin{equation}
\frac{\partial n}{\partial \mu} |U(0,q)|\sim
\begin{cases}
1, & q_0\ll q,\\
(\xi_0/l)^{d-z} (q \xi_0)^{-d-\Delta_2}, & \xi_0^{-1}\ll q \ll q_0,\\
(\xi_0/l)^{d-z} (q \xi_0)^{-2}, & q_\varkappa \ll q \ll \xi_0^{-1},\\
(\varkappa/q)^{d-1}
, & 0<q\ll q_\varkappa.
\end{cases}
\label{eq7a}
\end{equation}
Here the scale $q_0=\xi_0^{-1}(\xi_0/l)^{(z-d)/(d+\Delta_2)}$ is given by the condition $P(q)q^2\sim M(q) q^2\sim 1$. The condition $P(q)q^2(\varkappa/q)^{d-1} \sim M(q)q^2(\varkappa/q)^{d-1} \sim 1$ determines the momentum scale
\begin{equation}
q_\varkappa=\xi_0^{-1}(\varkappa \xi_0)^{{(1-d)}/{(3-d)}}(\xi_0/l)^{{(d-z)}/{(3-d)}}.
\label{eq8}
\end{equation}
We see that at the lowest momenta, $q\ll q_\varkappa$, in $d<3$ the interaction remains unscreened.
In $d=3$, this scale disappears, $q_\varkappa=0$. Thus, in $d=3$ the screened interaction remains long-ranged on scales larger than the localization length:
\begin{equation}
U(r)\sim \Delta \frac{\xi_0}{r}, \qquad r\gg \xi_0 .
\label{eqUR}
\end{equation}

We stress that in order to obtain the static screening in the localized state, one should use the dynamically
screened interaction. This is because the diffusion coefficient is itself proportional to $\omega$, so that
the frequency cancels out in Eq.~(\ref{eqUe}). Therefore, for the analysis of the static screening
it is incorrect first to set $-i \omega = 0$ and then cancel out the terms $D(\omega,q)q^2$ in
the ``static'' polarization operator in Eq.~(\ref{eqUe}).
Such a replacement, $\Pi(\omega=0,q) \to \partial n/\partial \mu,$ would yield incorrectly a conventional metallic screening $U(0,q)\sim e^2/(q^{d-1}+\varkappa^{d-1})$.
We see that the dynamically screened interaction is in fact static for $\omega\ll \Delta$.

As follows from Eq. \eqref{eq7a}, the effective dielectric function at $\omega \ll \Delta$ can be written as
\begin{equation}
\varepsilon(0,q) \sim
\begin{cases}
(\varkappa/q)^{d-1}, & q_0\ll q,\\
(\varkappa l)^{d-1} (q l)^{1+\Delta_2} (\xi_0/l)^{z+\Delta_2}, & \xi_0^{-1}\ll q \ll q_0,\\
(\varkappa l)^{d-1}(\xi_0/l)^{z+2-d} (q l)^{3-d}, & q_\varkappa \ll q \ll \xi_0^{-1},\\
1
, & 0<q\ll q_\varkappa.
\end{cases}
\label{eq7b}
\end{equation}
In particular, Eq. \eqref{eq7b} implies that in $d=3$ the effective permittivity ($\varepsilon$) and electric susceptibility ($\chi$), related via $\varepsilon = 1+4\pi \chi$, that determine the behavior of the Coulomb interaction at large distances $r\gg \xi_0$ [see Eq. \eqref{eqUR}], are divergent upon
 approaching the MIT according to $\varepsilon \sim \chi \sim (\xi_0/l)^{z-1} \gg 1$, in agreement with Ref. [\onlinecite{finkelstein90}].

\begin{figure}[t]
\centerline{\includegraphics[width=0.8\columnwidth]{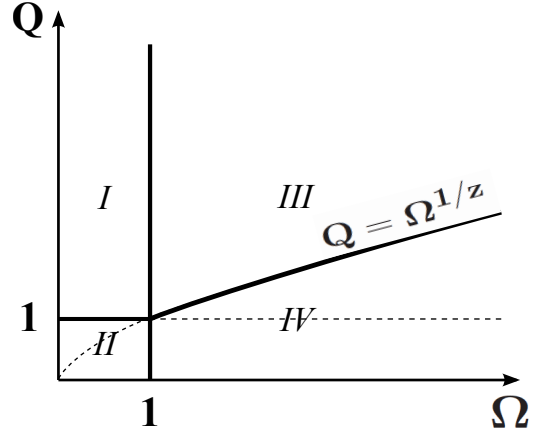}}
\caption{Sketch of different regions for the asymptotic behavior of the scaling functions $\mathcal{R}_D(\Omega,Q)$ and $\mathcal{R}_Z(\Omega,Q)$.}
\label{Fig_RDRZ}
\end{figure}

\section{Disorder-averaged tunneling density of states \label{AppTDOS}}

\begin{table*}[t]
\caption{The asymptotic behavior of the scaling function $\mathcal{R}_D(\Omega,Q)$. Here $c$ is a constant of the order unity, the exponent $\alpha$ is equal to $2$ for the Mott's formula and $1$ in the case of modifications of Ref.
\protect[\onlinecite{ES, ES1981, ESred}] due to Coulomb energy associated with the pairs of states involved in the transport.}
\label{Tab_IntDiffD}
\begin{tabular}{l|lc|lc}
& & $0\leqslant \Omega\ll 1$ & & $1\ll \Omega$ \\
\hline $\max\{\Omega^{1/z}, 1\} \ll Q$ & (I): &
$Q^{d-2+\Delta_2}(-i\Omega + c |\Omega|^\alpha)$
 & (III): &  $Q^{d-2+\Delta_2} |\Omega|^{-\Delta_2/z}$ \\
$0\leqslant Q \ll \max\{\Omega^{1/z}, 1\}$& (II): &  $-i\Omega + c |\Omega|^\alpha$& (IV): & $|\Omega|^{(d-2)/z}$\\
\end{tabular}
\end{table*}

In this appendix we analyze the disorder-averaged tunneling density of states (TDOS) at the insulating side ($g<g_*$) of the MIT in $d=2+\epsilon$. At $T=0$,
it is convenient to rewrite Eq. \eqref{e3} in the equivalent form by using the real-time representation 
derived in Ref. [\onlinecite{kamenev99}]:
\begin{equation}
\langle\rho(E)\rangle=\frac{\rho_0}{\pi} \int\limits_{-\infty}^\infty dt \frac{\sin(|E|t)}{t}
\exp\bigl[-J_c(t)\bigr] \cos\bigl[J_s(t) \bigr].
   \label{KA-DOS}
\end{equation}
Here $\rho_0 = 1/(E_0 l^d)$ is the ultraviolet value of the DOS  which differs  from $\partial n/\partial \mu$ by the Fermi-liquid ballistic renormalizations (we shall disregard this difference below) and
\begin{align}
J_c(t) & = \int\limits_0^\infty \frac{d\omega}{\pi} \int \frac{d^dq}{(2\pi)^d} \im \mathcal{V}^R(\omega,q)\ (1-\cos\omega t), \notag \\
J_s(t) & = \int\limits_0^\infty \frac{d\omega}{\pi}  \int \frac{d^dq}{(2\pi)^d}\ \text{Im} \mathcal{V}^R(\omega,q)\ \sin\omega t, \label{Js}
\end{align}
with the retarded propagator
\begin{equation}
\mathcal{V}^R(\omega,q)=\frac{Z U_0(q)}{[D q^2-i Z \omega][D q^2(1+(\partial n/\partial \mu) U_0(q))-i\omega]}.
\label{calV}
\end{equation}
We remind the reader that the energy $E$ is measured with respect to the chemical potential $\mu$.
With the help of Eq. \eqref{eqIneq}, Eq. \eqref{calV} can be simplified:
\begin{equation}
\mathcal{V}^R(\omega,q)\simeq \frac{1}{\rho_0}\frac{Z}{[Dq^2-i Z\omega][D q^2-i\omega (q/\varkappa)^{d-1}]}.
\label{calV1}
\end{equation}
We mention that at $(q/\varkappa) \to 0$, the expression \eqref{calV1} for $\mathcal{V}^R(\omega,q)$ corresponds to Eq. (10) from the main text for its Matsubara counterpart $\mathcal{V}(i\omega_m,q)$.

The diffusion coefficient $D(\omega,q)$ (see Eq.\eqref{eqDcInt}) and Finkelstein's frequency renormalization factor $Z(\omega,q)$ parametrize the interacting {\it mesoscopic} diffuson (we omit the dephasing rate induced due to interaction)
\begin{equation}
\mathcal{D}^{\rm int}(\omega,q) = \frac{1}{D(\omega,q)q^2-i Z(\omega,q)\omega} .
\label{IntDiffuson}
\end{equation}
The frequency renormalization factor can be written in the following scaling form: \cite{finkelstein90}
\begin{equation}
Z(\omega,q) = (\xi_0/l)^{z-d} \mathcal{R}_Z(\omega/\Delta, q\xi_0) .
\end{equation}

The asymptotic behavior of the scaling functions  $\mathcal{R}_D(\Omega,Q)$ and $\mathcal{R}_Z(\Omega,Q)$ in different domains of the $Q=q\xi_0$ and $\Omega=\omega/\Delta$ plane (see Fig. \ref{Fig_RDRZ}) is summarized in Tables \ref{Tab_IntDiffD} and \ref{Tab_IntDiffZ}. In regions (I) and (III) we include possible effect of the multifractality in the presence of the Coulomb interaction [\onlinecite{burmistrov13}] (see also [\onlinecite{amini13}]) on the diffusion coefficient which is characterized by the interacting multifractal exponent $\Delta_2<0$ (cf. Eqs. \eqref{eq6} and \eqref{eq6a}). 
It is worth noting that, in view of the gauge invariance (preserved by the energy-independent quantities like the polarization operator),
the effect of the multifractality on the diffusion coefficient in the case of Coulomb interaction deserves a separate detailed study. As we shall demonstrate below, even if multifractality affects the diffusion coefficient in the case of Coulomb interaction, it does not influence the scaling results for the disorder-averaged TDOS.

\begin{table}[b]
\caption{The asymptotic behavior of the scaling function $\mathcal{R}_Z(\Omega,Q)$.}
\label{Tab_IntDiffZ}
\begin{tabular}{l|lc|lc}
& & $0\leqslant \Omega\ll 1$ & & $1\ll \Omega$ \\
\hline $\max\{\Omega^{1/z}, 1\} \ll Q$ & (I): &
$Q^{d-z}$
 & (III): &  $Q^{d-z} $ \\
$0\leqslant Q \ll \max\{\Omega^{1/z}, 1\}$& (II): &  $1$& (IV): & $|\Omega|^{(d-z)/z}$\\
\end{tabular}
\end{table}

It is convenient to rewrite Eq. \eqref{calV1} in terms of the scaling functions $\mathcal{R}_{D/Z}(\Omega,Q)$ and dimensionless variables $\Omega$ and $Q$. Provided condition $z+\Delta_2>0$ is fulfilled, one finds with the help of Eq. \eqref{eqIneq}:
\begin{equation}
\mathcal{V}^R(\omega,q) =  \frac{{\xi_0^{d}}/{\Delta}}{Q^2 \mathcal{R}_D(\Omega,Q)}\
\frac{1}{Q^2 \mathcal{R}_D(\Omega,Q)/\mathcal{R}_Z(\Omega,Q)-i\Omega}
.
\label{calV2}
\end{equation}

We start the analysis of the momentum integral in Eq. \eqref{Js} from the case $\Omega \gg 1$. 
In the region (III), if the inequality $z+3\Delta_2/2>0$ is fulfilled, the integral is dominated by momenta $Q\sim \Omega^{1/z}$, and we find
\begin{gather}
\int\limits_{\Omega^{1/z}\lesssim  Q} \frac{d^d q}{(2\pi)^d}\im \mathcal{V}^R(\omega,q)
= \frac{S_d}{(2\pi)^d}  \frac{\Omega}{\Delta} \int\limits_{\sim \Omega^{1/z}}^\infty dQ \ \frac{Q^{d-1}\mathcal{R}_Z^2}{\mathcal{R}^3_D Q^6} \notag \\
\simeq  \frac{S_d\Omega^{(z+d-6)/z}\mathcal{R}_Z^2(\Omega,\Omega^{1/z})}{(2z+3\Delta_2)(2\pi)^d\Delta\mathcal{R}^3_D(\Omega,\Omega^{1/z})}.
\label{eqIntIII}
\end{gather}
Here, $S_d = 2\pi^{d/2}/\Gamma(d/2)$ is the area of the $d$-dimensional sphere. In the region (IV),
we obtain
\begin{gather}
\int\limits_{Q\lesssim \Omega^{1/z}} \frac{d^d q}{(2\pi)^d}\im \mathcal{V}^R(\omega,q)
= \frac{S_d}{(2\pi)^d\Delta \Omega} \int\limits_0^{\sim \Omega^{1/z}} dQ \ \frac{Q^{d-1}}{\mathcal{R}_D Q^2} \notag \\
= \frac{1}{2\pi\Delta \Omega \mathcal{R}_D(\Omega,0)}\Bigl [ \frac{1}{\epsilon}+O(1)\Bigr ] .
\label{eqIntV1}
\end{gather}
Comparing Eqs. \eqref{eqIntIII} and \eqref{eqIntV1}, we find that in $d=2+\epsilon$ dimensions for $\Omega \gg 1$, if the inequality $z+3\Delta_2/2>0$ holds, the main contribution to the integral over momenta comes from the domain $Q\ll 1$ in the region (IV).

In the opposite case of $\Omega\ll 1$, the integral over momenta is dominated by the contribution from the region (II) if the inequality $z+2\Delta_2>0$ is fulfilled. Then, we find
 \begin{gather}
\int\limits_{0<Q\lesssim 1} \frac{d^d q}{(2\pi)^d}\im \mathcal{V}^R(\omega,q)
= \re \frac{S_d/((2\pi)^d\Delta)}{\Omega+i 0^+} \int\limits_0^{\sim 1} dQ \ \frac{Q^{d-1}}{\mathcal{R}_D Q^2} \notag \\
= \frac{1}{2\pi\Delta} \re \frac{1}{(\Omega+i0^+)\mathcal{R}_D(\Omega,0)}\Bigl [ \frac{1}{\epsilon}+O(1)\Bigr ] .
\label{eqIntV2}
\end{gather}
Importantly, the leading contribution to this integral is proportional to $1/\epsilon$, similarly to the case of critical frequency domain,
$\omega\gg \Delta.$ We mention that Eqs. \eqref{eqIntV1} and \eqref{eqIntV2} can be written in the unified way if one takes into account that in $d=2+\epsilon$ the dimensionless conductance $g(\omega) = 2\pi \mathcal{R}_D(\Omega,0)$:
\begin{equation}
\int \frac{d^d q}{(2\pi)^d}\im \mathcal{V}^R(\omega,q)  = \re \frac{1}{\epsilon} \frac{1}{(\omega+i0^+)g(\omega)} .
\label{eqIntV3}
\end{equation}
Here we neglect terms which are finite at $\epsilon\to 0$.

As it was demonstrated above, even if the multifractality affects the diffusion coefficient in the case of Coulomb interaction, it does not influence the integral \eqref{eqIntV3} over momenta for $\omega\gg \Delta$ ($\omega\ll \Delta$) if the inequality $z+3\Delta_2/2>0$ ($z+2\Delta_2>0$) holds. Below we shall demonstrate
that the disorder-averaged TDOS at $|E|\gg\Delta$ is determined by the integral \eqref{eqIntV3}
with $\omega\gg \Delta$. Therefore, provided $z+3\Delta_2/2>0$, the possible multifractality in the diffusion coefficient is not important for the disorder-averaged TDOS at $|E|\gg\Delta$. Moreover, the possible multifractality contribution does not affect Eq.~(\ref{KA-DOS}) also at $|E|\ll \Delta$,
if the condition $z+2\Delta_2>0$ is met.

In $d=2+\epsilon$ dimensions, for the MIT in a system of disordered electrons with Coulomb interaction with broken time-reversal and spin-rotational symmetries due to the presence of magnetic impurities (the ``MI(LR)" class in terminology of Belitz and Kirkpatrick, \cite{belitz94} the dynamical exponent $z$ is known up to the second loop order \cite{baranov99,baranov02}
\begin{equation}
z=2+\frac{\epsilon}{2}+\left(2A-\frac{\pi^2}{6}-3\right)\frac{\epsilon^2}{4}+ \mathcal{O}(\epsilon^3),
\label{eqz_d}
\end{equation}
where 
\begin{align}
A & =\frac{1}{16}\Biggl [\frac{139}{6}+\frac{(\pi^2-18)^2}{12}+\frac{19}{2}\zeta (3)+\Bigl ( 16 + \frac{\pi ^2}{3} \Bigr )\ln ^{2}2  \notag \\
& - \Bigl (44-\frac{\pi ^{2}}{2}+7\zeta (3)\Bigr ) \ln 2+16\mathcal{G}-\frac{1}{3}\ln ^{4}2-8\lit\left(\frac{1}{2}\right)\Biggr ] \notag \\
& \approx 1.64 .
\end{align}
Here $\mathcal{G} \approx 0.915 $ denotes the Catalan constant, $\zeta(x)$ stands for the Riemann zeta-function, and $\lit(x) = \sum_{k=1}^\infty x^k/k^4$ denotes the polylogarithm. Recently, the multifractal exponent $\Delta_2$ was also computed up to the second loop order: \cite{burmistrov13}
\begin{equation}
\Delta_2  =
-\frac{\epsilon}{2}\Bigl [1 + \left (1-A-\frac{\pi^2}{12}\right ) \epsilon \Bigr ]+O(\epsilon^3) .
\label{eqd2_d}
\end{equation}
Therefore, for the MIT in the class ``MI(LR)" in $d=2+\epsilon$ dimensions the inequality $z+2\Delta_2>0$ holds.

With the help of Eq. \eqref{eqIntV3}, the functions $J_{c}(t)$ and $J_{s}(t)$ (see Eq. \eqref{Js}) can be written as
\begin{align}
\begin{split}
J_c(t) & =\frac{t^2}{8\epsilon}\lim \limits_{\omega\to 0} \im \frac{\omega^2}{g(\omega)} +  {\rm p.v.\,} \int\limits_0^{\infty} \frac{d\omega}{\epsilon \pi}  \frac{1-\cos\omega t}{\omega} \re \frac{1}{g(\omega)} ,  \\
J_s(t) & = \frac{t}{4\epsilon}\lim \limits_{\omega\to 0} \im \frac{\omega}{g(\omega)} +  {\rm p.v.\,} \int\limits_0^\infty \frac{d\omega}{\epsilon \pi}  \frac{\sin\omega t}{\omega} \re \frac{1}{g(\omega)} .
\end{split}\label{Jsc1}
\end{align}
In order to make analytical estimates for the disorder-averaged TDOS in the insulating side ($g<g_*$) of the metal-insulator transition we assume that at $\omega>\Delta$ the real part of the conductance $g(\omega)$ can be written as $\re g(\omega) = g_* f_g(\omega/\Delta)$ where the dependence on $\omega$ is due to the frequency induced length $L_\omega \sim |\omega|^{-1/z}$. Since at $\omega\gg \Delta$, the renormalization of the conductance up to the scale $L_\omega$ is governed by the interacting critical point, the function $f_g(x)$ has the following asymptote at $x\gg 1$: $f_g(x) = 1-x^{-1/(z\nu)}$. At $\omega<\Delta$, we use the scaling form for the diffusion coefficient in the region (II) (see Table \ref{Tab_IntDiffD}): $g(\omega) = -i\omega/\Delta +c (|\omega|/\Delta)^\alpha$ with $1 \leqslant\alpha \leqslant 2$.  Then, we find
\begin{align}
J_c(t)  = & \int\limits_0^1 \frac{dx}{\pi \epsilon} \frac{1-\cos(\Delta t x)}{x^{3-\alpha}} \frac{c}{1+c^2 x^{2\alpha-2}}  \notag \\
& + \int\limits_1^{E_0/\Delta}  \frac{dx}{\pi \epsilon g_*} \frac{1-\cos(\Delta t x)}{x f_g(x)}
 ,   \notag\\
J_s(t)  = & \frac{c(\alpha) \Delta t}{4\epsilon} + \int\limits_0^1 \frac{dx}{\pi \epsilon} \frac{\sin(\Delta t x)}{x^{3-\alpha}}\frac{c}{1+c^2 x^{2\alpha-2}} \notag \\
& +\int\limits_1^{\infty}  \frac{dx}{\pi \epsilon g_*} \frac{\sin(\Delta t x)}{x f_g(x)} ,
\label{eqJsJcint}
\end{align}
where $c(\alpha) = 1$ for $1< \alpha \leqslant 2$ and $c(1) = 1/(1+c^2)$. 
At interacting criticality, $|E|\gg \Delta$, the integral in Eq. \eqref{KA-DOS} is determined by small values of $t$,
$t \ll 1/\Delta$. At $t \to 0$, behavior of functions $J_c(t)$ and $J_s(t)$ is determined by the integrals over $x>1$ in Eqs. \eqref{eqJsJcint}. We obtain the following asymptotic expressions
\begin{equation}
J_c(t) = \frac{1}{\pi \epsilon g_*}
\ln E_0 t, \quad J_s(t) = \frac{1}{2 \epsilon g_*} , \quad \Delta t \ll e^{-z\nu} .
\label{eqAe1}
\end{equation}
We mention that the results \eqref{eqAe1} are not sensitive to the precise form of the function $f_g(x)$. By using asymptotic expressions \eqref{eqAe1},
we find from Eq. \eqref{KA-DOS} the following result for the disorder-averaged TDOS:
\begin{equation}
\langle\rho(E)\rangle \sim \rho_0 \left (\frac{|E|}{E_0}\right )^\beta , \qquad \beta = \frac{1}{\pi g_* \epsilon}, \qquad
|E| \gg\Delta e^{z\nu}.
\label{eqTDOSc}
\end{equation}
In $d=2+\epsilon$ dimensions, the critical point of the MIT point in the class ``MI(LR)" is also known up to the second loop [\onlinecite{baranov02}]
\begin{equation}
\frac{1}{\pi g_*} = \frac{\epsilon}{2}(1-A\epsilon)+O(\epsilon^3) .
\label{eqg*d}
\end{equation}
Therefore, in $d=2+\epsilon$ dimensions for the class ``MI(LR)" the exponent $\beta$ which determines the power-law behavior of the disorder-averaged TDOS is given as \cite{finkelstein90,belitz94}
\begin{equation}
\beta = \frac{1}{2} + O(\epsilon) .
\label{eqbetad}
\end{equation}
Note that in order to find the exponent $\beta$ at the order $\epsilon$ one needs to compute Eq. \eqref{eqIntV3} in the next order in $\epsilon$.

At energies $\Delta \ll |E| \ll \Delta\, e^{z\nu}$ the integral in Eq. \eqref{KA-DOS} is dominated by the range
$\Delta^{-1}\exp(-z\nu) \ll t \ll \Delta^{-1}$. At $\exp(-z\nu) \ll \Delta t \ll 1$ the asymptotic behavior of functions $J_c(t)$ and $J_s(t)$ is as follows
\begin{align}
J_c(t) & = \frac{1}{\pi \epsilon g_*}
\ln \frac{E_0}{\Delta}  + \frac{z\nu}{\pi \epsilon g_*}
\ln \frac{z\nu}{\ln [\lambda/(\Delta t)]}, \notag \\
 J_s(t) &= \frac{1}{2 \epsilon g_*} + \frac{c(\alpha)\Delta t}{4\epsilon}, \quad e^{-z\nu} \ll \Delta t \ll 1.
\label{eqAe1_2}
\end{align}
Here $\lambda=\exp[(e-1)z\nu]$ is determined from the coincidence of asymptotes \eqref{eqAe1} and \eqref{eqAe1_2} for $J_c(t)$ at $t=\exp(-z\nu)/\Delta$. At $\Delta/\epsilon \ll |E| \ll \Delta\, e^{z\nu}$, we can neglect the second term in the asymptotic expression \eqref{eqAe1_2} for $J_s(t)$. Then performing integration over $t$ in Eq.\eqref{KA-DOS} we find for $\Delta/\epsilon \ll |E| \ll \Delta\, e^{z\nu}$ that
\begin{equation}
\langle\rho(E)\rangle \sim \rho_0 \left (\frac{\Delta}{E_0}\right )^\beta \left (
\ln \frac{z\nu}{\ln (\lambda |E|/\Delta)} \right )^{- z\nu\beta} .
\label{eqrhoL_2}
\end{equation}
We emphasize that the results \eqref{eqTDOSc} and \eqref{eqrhoL_2} are determined by the behavior of functions $J_c(t)$ and $J_s(t)$ that comes from the frequencies $\omega>\Delta$ in the integrals of Eq. \eqref{eqJsJcint}. 

In the region $\Delta \ll |E| \ll \Delta/\epsilon$, the second term in the asymptotic expression for $J_s(t)$ in Eq.\eqref{eqAe1_2} restricts the integral in Eq.\eqref{KA-DOS} to the domain $t\lesssim \epsilon$. Hence, for 
$\Delta \ll |E| \ll \Delta/\epsilon$ we obtain
\begin{equation}
\langle\rho(E)\rangle \sim \rho_0 \left (\frac{\Delta}{E_0}\right )^\beta \left (
\ln \frac{z\nu}{\ln [\lambda/(\epsilon\Delta)]} \right )^{- z\nu\beta} \frac{\epsilon |E|}{\Delta} .
\label{eqrhoL_3}
\end{equation}

We have shown in the main text, that the critical behavior of the TDOS, Eq.~(\ref{eqTDOSc}),
at $|E|\sim \Delta$ crosses over to the Coulomb-gap behavior. At $|E|\ll \Delta$ the leading contribution to the functions $J_c(t)$ and $J_s(t)$ is also proportional to $1/\epsilon$ (see Eq. \eqref{eqIntV2}), similarly to the case 
$\omega\gg \Delta$. Therefore, it is instructive to evaluate Eq.~\eqref{KA-DOS} also at the lowest energies by using Eq. \eqref{eqIntV3} and the scaling form of the diffusion
 coefficient \eqref{eqDcInt} as presented in Appendix \ref{Sec1}.
For $|E|\ll \Delta$ the integral in Eq. \eqref{KA-DOS} is determined by large values of $t$,
$t \gg 1/\Delta$. We find the following asymptotic expressions at $\Delta t \gg 1$
\begin{align}
J_c(t) & =\frac{1}{\pi \epsilon g_*}
 \ln \frac{E_0}{\Delta} + \frac{c}{2-\alpha} \frac{(\Delta t)^{2-\alpha}}{\pi \epsilon}
, \notag \\
J_s(t)  & =  \frac{c(\alpha) \Delta t}{4\epsilon}+\frac{c}{\alpha-1}\frac{(\Delta t)^{2-\alpha}}{\pi \epsilon} .
\label{eqJcJstL}
\end{align}
We mention that in the case $\alpha=2$ the leading contribution to asymptote of $J_c(t)$ at large $t$
is logarithmic. But, in order to find the power of the logarithm one needs to take into the logarithms in the Mott formula. A similar problem occurs with the large-$t$ asymptote of $J_s(t)$ for $\alpha=1$. As one can see, the asymptotic expressions for $J_c(t)$ and $J_s(t)$ are sensitive to the precise form of the frequency dependence of the diffusion coefficient at $\omega\ll \Delta$. Since the scaling function $\mathcal{R}_D(\Omega,Q)$ remains to be calculated from the microscopic theory in the case of Coulomb interaction, in what follows we do not use the precise form of $t$ dependence in Eq. \eqref{eqJcJstL}.

Fortunately, in order to find
the energy dependence of the disorder-averaged TDOS at $|E|\ll\Delta$, we do not need to known the precise asymptotic form of $J_c(t)$ at $t\gg 1/\Delta$. Provided the function $J_c(t)$ grows faster than the first power of $\ln (\Delta t)$, we obtain the linear dependence of the disorder-averaged TDOS on energy.
Since $\langle\rho(E)\rangle$ are continuos function we obtain that its behavior at $|E|\ll \Delta$ is still given by Eq. \eqref{eqrhoL_3}.
If the large-$t$ asymptotic behavior of the function $J_c(t)$ is logarithmic, then  at $|E|\ll \Delta$ the energy dependence of the disorder-averaged TDOS will be a power law with some non-trivial exponent. It is worthwhile to mention that if,  as in the main text, one neglects the real part of the conductance $g(\omega)$ for $\omega<\Delta$ and neglects contributions to the functions $J_{c,s}(t)$ from $\omega>\Delta$, then the disorder-averaged TDOS will vanish for $|E|<\Delta/(4\epsilon)$. We mention that in the main text, $\Delta$ is defined to be equal to the half of the hard gap and, therefore, is different by a factor $4\epsilon$ from $\Delta$ used here.  As discussed in the main text, an uncorrelated averaging over positions of such local hard gaps yields a linear behavior of $\langle\rho(E)\rangle$ similar to Eq. \eqref{eqrhoL_3}.
We remind that the above calculation at $|E| \alt \Delta$ [$g(L_E)\alt 1$] yields only the upper bound of the TDOS. 


\section{Dephasing of the single-particle excitations above $E_c$: Golden-Rule calculations \label{AppDep}}

In this appendix we present the derivation of the result for the exponent $z_\phi^{\rm n}$. We start with the conventional formula for the inelastic-scattering (dephasing) rate for a single-particle excitation
at energy $E$ expressed in terms of
diffusion propagators and (dynamically screened) Coulomb interaction $U(\omega,q)$:
\begin{align}
\frac{1}{\tau_\phi(E)}& \sim \rho_0 \int d\omega dE^\prime  d^dq 
\re \mathcal{D}(\omega,q;E) \re \mathcal{D}(\omega,q;E^\prime) \notag \\
& \times |U(\omega,q)|^2 [1-f(E-\omega)] f(E^\prime) [1-f(E^\prime+\omega)].
\label{eq1}
\end{align}
Here
$\mathcal{D}(\omega,q;E)$ is the diffusion propagator at energy $E$ (measured from the chemical potential $\mu$)
and $f(E)$ is the distribution function for single-particle states.

We are interested in the zero-temperature dephasing of a single-particle critical state with energy $E>E_c$ near the
(renormalized by the interaction) single-particle
mobility edge $E_c$: $(E-E_c)/E_c \ll 1$. The dephasing occurs due to transitions between delocalized states with energies $E>E_c$ and $E-\omega>E_c$ accompanied by the excitation of electron-hole pairs near the chemical potential. At $T=0$ we thus have the following conditions which restrict the phase space for inelastic scattering:
\begin{equation}
0<\omega<E-E_c, \qquad E^\prime<0, \qquad E^\prime+\omega>0.
\label{eq3}
\end{equation}
Under these conditions, the product of distribution functions in Eq.~\eqref{eq1} is equal to unity.

We remind that in $d=2+\epsilon$ the single-particle mobility edge $E_c$ is much higher than the characteristic energy $\Delta\sim \xi_0^{-z}$ whereas in $d=3$ the two scales $E_c$ and $\Delta$ coincide. The particle-hole excitations are localized for energies smaller than $\Delta$.
Therefore, in order to find the critical exponent of the dephasing length, we consider the energies
satisfying: $E-E_c\ll \Delta$, such that the particle-hole excitations with $\omega < E-E_c$ are localized and described by the ``localized'' interacting diffusion propagator.  We note that this diffuson corresponds to the irreducible polarization operator (it can be considered as a part of the screening) and hence the frequency in this diffuson is not renormalized by the Finkelstein's frequency renormalization factor $Z$, in contrast to the mesoscopic diffuson (see Appendix \ref{Sec1}).

Furthermore, we assume that both critical noninteracting ($c$) and localized interacting ($l$) diffusons are energy-independent
\begin{equation}
\mathcal{D}(\omega,q;E) \equiv \mathcal{D}_c(\omega,q), \quad \mathcal{D}(\omega,q;E^\prime) \equiv \mathcal{D}_l(\omega,q).
\label{eq3a}
\end{equation}
In view of Eq.~(\ref{eq3}), the integration over $E^\prime$  then yields just $\omega$. Representing
the structure $\re\mathcal{D}_1\re\mathcal{D}_2|U|^2$ equivalently through $\re\mathcal{D}_1 \im U$,
we arrive at
\begin{equation}
\frac{1}{\tau_\phi(E)} \sim  - \rho_0 \int\limits_0^{E-E_c} d\omega  \int\limits_0^{1/\xi_c} dq\ q^{d-1} \re \mathcal{D}_c(\omega,q) \im U(\omega,q).
\label{eq4}
\end{equation}
Here the integration over the transferred momentum is restricted by $q<\xi_c^{-1}$ which is the ultraviolet
cutoff of the renormalized theory.

It is worth emphasizing, that Eq.~(\ref{eq1}) describes the Hartree contribution to the dephasing rate.
For short-ranged interaction, one encounters a strong Hartree-Fock cancellation in the critical regime. \cite{LeeWang,Wang2,AnnPhys2011}
However, in our case of long-ranged Coulomb interaction, the exchange counterparts of Eq.~(\ref{eq1}) are subleading.
Indeed, the characteristic scale of the localized interacting diffuson is $\xi_c$ which is the ``ballistic scale'' for the critical noninteracting diffuson. As we will see, the dominant contribution to the dephasing rate comes from much larger scales related
to $E-E_c \ll \Delta$. On such scales the Hartree contribution dominates over exchange term. In the latter one interaction line necessarily connects the two point separated by distance $\alt \xi_c$, whereas in the former there is no such restriction.

By using Eqs~\eqref{eqUe}, \eqref{eq10}, \eqref{eq6}, and \eqref{eq11a}, the imaginary part of the screened Coulomb interaction involved in Eq.~\eqref{eq4} can be written as
\begin{align}
\im U(\omega,q)
&  \sim - \frac{1}{\rho_0}\left(\frac{\xi_0}{l}\right)^{d-z} (q\xi_0)^{-2} \notag \\
& \times
\begin{cases}
(q/q_\varkappa)^{2(3-d)}, & \quad 0<q\leqslant q_\varkappa ,\\
1,  & \quad q_\varkappa< q \leqslant \xi_0^{-1} , \\
(q\xi_0)^{-d+2-\Delta_2}, & \quad \xi^{-1}_0 <q .
\end{cases}
\label{eq13}
\end{align}
Here we consider the case of the Mott's law modified
by Coulomb energy associated with the pairs of states involved in the transport \cite{ES, ES1981, ESred} (i.e. $\alpha=1$) and neglect logarithmic factors in the real part of {\it ac} conductivity. Substituting  Eqs~\eqref{eq6}, \eqref{eq11a}, and (\ref{eq13}) into Eq.~(\ref{eq4}), we find
\begin{align}
\frac{1}{\tau_\phi(E)} & \sim \xi_0^{d-2} \Delta 
\int\limits_0^{E-E_c} d\omega  \Biggl [ \int\limits_{q_\varkappa}^{1/\xi_0} dq\
q^{d-3} \re \mathcal{D}_c(\omega,q) \notag \\
& +
 \int\limits_0^{q_\varkappa} dq\
\left (\frac{q}{q_\varkappa^2}\right)^{3-d} \re \mathcal{D}_c(\omega,q) \notag \\
& +
  \int\limits_{1/\xi_0}^{1/\xi_c} dq\
q^{d-3} (q\xi_0)^{-d+2-\Delta_2}\re \mathcal{D}_c(\omega,q)\Biggr ]
.
\label{eq14b}
\end{align}
In the experimentally relevant case $d=3$ the second (that of $0<q<q_\varkappa$) and  third (that of $1/\xi_0<q<1/\xi_c$) contributions disappear (since $q_\varkappa=0$ and $\xi_c \sim \xi_0$)
and we obtain
\begin{equation}
\frac{1}{\tau_\phi(E)} \sim \xi_0 \Delta
\int\limits_0^{E-E_c} d\omega  \int\limits_{0}^{1/\xi_0} dq\
\re \mathcal{D}_c(\omega,q).
\label{eq14}
\end{equation}
\begin{figure}[b]
\centerline{\includegraphics[width=0.8\columnwidth]{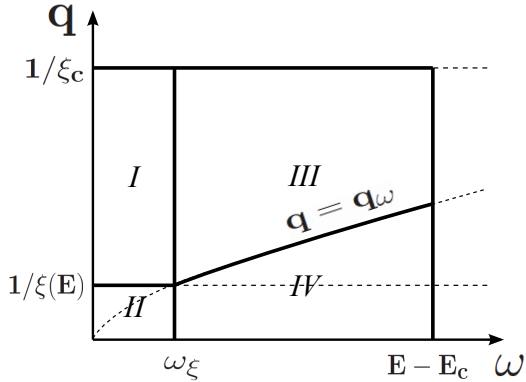}}
\caption{Sketch of different regions for behavior of the diffusion coefficient $D_{\rm n}(\omega,q)$ in the critical noninteracting diffuson $\mathcal{D}_c(\omega,q)$.}
\label{fig1D}
\end{figure}

\begin{table*}[t]
\caption{Diffusion coefficient $D_{\rm n}(\omega,q)$ of the critical noninteracting diffuson $\mathcal{D}_c(\omega,q)$.}
\label{Tab_Dc}
\begin{tabular}{l|lc|lc}
& & $0\leqslant \omega<\omega_\xi$ & & $\omega_\xi \leqslant \omega < E-E_c$ \\
\hline $\max\{q_\omega, 1/\xi(E)\} \leqslant q < \xi^{-1}_c$ & I: &
$D_c (q\xi_c)^{d-2} (q\xi(E))^{\Delta_2^{\rm n}}$
 & III: &  $D_c (q\xi_c)^{d-2} (q/q_\omega)^{\Delta_2^{\rm n}}$ \\
$0\leqslant q < \max\{q_\omega, 1/\xi(E)\}$& II: &  $D_c(\xi_c/\xi(E))^{d-2}$& IV: & $D_c(q_\omega \xi_c)^{d-2}$\\
\end{tabular}
\end{table*}

Since we are interested in the dephasing length that cuts off the scaling of the conductance or the mesoscopic fluctuations in the noninteracting RG, the critical noninteracting diffuson
\begin{equation}
\mathcal{D}_c(\omega,q) = \frac{1}{D_{\rm n}(\omega,q)q^2-i\omega} .
\end{equation}
is understood as the mesoscopic diffuson at scales $\lesssim \xi_c$ where the interacting RG was operated.
Therefore, the diffusion constant of the critical noninteracting diffuson $\mathcal{D}_c(\omega,q)$ at the length scale $\xi_c$ can be estimated as $D_c = D(E_c,\xi_c^{-1})/Z(E_c,\xi_c^{-1}) \sim g_*^{\rm n} E_c \xi_c^2$.
As long as the behavior of the critical diffuson is concerned, we have four different domains in the $\omega$ -- $q$ plane (see Fig.~\ref{fig1D}) for behavior of the diffusion coefficient  (see Table~\ref{Tab_Dc}). At given frequency, the domains are separated either by the momentum scale $q_\omega = \xi_c^{-1}(\omega \xi_c^2/D_c)^{1/d}$ or by the inverse localization length $\xi(E) = \xi_c (E/E_c-1)^{-\nu_{\rm n}}$ for the noninteracting Anderson transition (see Fig.~\ref{fig1D}). Regions of small and large frequencies are separated by the energy scale $\omega_\xi = (D_c/\xi_c^2) (\xi(E)/\xi_c)^{-d}$ (see Fig.~\ref{fig1D}). The condition $\nu_{\rm n} d >1$ implies that $\omega_\xi \ll E-E_c$.
In $d=2+\epsilon$ dimensions for the Anderson transition in the Wigner-Dyson class A (which is noninteracting counterpart of the interacting ``MI(LR)" class) the exponent of the localization length is known up to the five-loop order, $\nu_{\rm n} = 1/(2\epsilon)-3/4$. \cite{HikamiWegnerB} Therefore, the inequality $\nu_{\rm n} d >1$ is fulfilled. We mention that in the domains (I) and (III) the diffusion coefficient $D_{\rm n}(\omega,q)$ is depends on the multifractal exponent $\Delta_2^{\rm n}<0$ for the noninteracting Anderson transition. In $d=2+\epsilon$ dimensions for the Anderson transition in the Wigner-Dyson class A, the multifractal exponent is known up to the forth loop $\Delta_2^{\rm n} = - (2\epsilon)^{1/2} - 3\zeta(3) \epsilon^2/8$. \cite{Wegner}

Substituting $\re \mathcal{D}_c(\omega,q)$ into Eq.~(\ref{eq14b}), we observe
that the first two integrals over momenta are dominated by $q\sim q_\omega$ (the domains (III) and (IV)). The contribution of the region
$\xi_0^{-1}<q<\xi_c^{-1}$ will be discussed. Let us first assume that $q_\varkappa \ll q_{E-E_c}=\xi_c^{-1}[(E-E_c) \xi_c^2/D_c]^{1/d} \ll \xi_0^{-1}$. Evaluating the integral over $q$ in the domain IV, we see that the frequency integral is dominated by the upper limit $\omega=E-E_c$. Then we find
\begin{gather}
\Biggl [ \frac{1}{\tau_\phi(E)} \Biggr ]_{\rm IV}  \sim \bigl( {\xi_c}/{\xi_0} \bigr)^{2-d} \bigl ( g_*^{\rm n}\bigr )^{2/d}
\int\limits_{\omega_\xi}^{E-E_c} d\omega\  \left(\frac{\omega}{E_c}\right)^{-2/d}  \notag \\
 \sim \frac{E_c}{\sqrt{\epsilon}} \left(\frac{E-E_c}{E_c}\right)^{1-2/d} , \notag \\
  g_*^{\rm n} (q_\varkappa \xi_c)^d  \ll \frac{E-E_c}{E_c} \ll \frac{\Delta}{E_c} .
\label{eq27}
\end{gather}
Here we have used that $(\xi_c/\xi_0)^{2-d}
\sim 1$ and $(g_*^{\rm n})^{2/d}
\sim 1/\sqrt{\epsilon}$ in $d=2+\epsilon$. The condition $q_{E-E_c}\gg q_\varkappa$ necessary breaks down
for energies sufficiently close the mobility edge.
For such energies the second contribution (from the region $0<q<q_\varkappa$) in Eq.~(\ref{eq14b}) becomes essential.
Provided  $-2<\Delta_2^{\rm n}<0$ the two distinct cases are possible (i) $4-2d-\Delta_2^{\rm n}<0$ and
(ii) $4-2d-\Delta_2^{\rm n}\geqslant 0$. We mention that in $d=2+\epsilon$ for the Anderson transition in the Wigner-Dyson class A the case (ii) is realized at small $\epsilon$. As we approach $d=3$ from below, we expect that the case (i) should be realized.

In the case (i) the dominant contribution to the dephasing rate
comes from momenta $q\sim q_\omega$, yielding
\begin{equation}
\frac{1}{\tau_\phi(E)}\sim E_c (q_\varkappa \xi_c)^{2(d-3)} \left(\frac{\xi_0}{\xi_c}\right)^{d-2-z}
\left(\frac{E-E_c}{g_*^{\rm n} E_c}\right)^{4/d-1},
\label{eq39a}
\end{equation}
and hence ($L_\phi \sim (\tau_\phi)^{1/d}$)
\begin{equation}
L_\phi \propto (E-E_c)^{-1/z^{\rm n}_\phi},\qquad  z^{\rm n}_\phi = {d^2}/({4-d}).
\label{eq54}
\end{equation}
In the case (ii) the main contribution to the dephasing rate comes from momenta $q\sim q_\varkappa$. Then, we obtain
\begin{equation}
\frac{1}{\tau_\phi(E)}\sim E_c 
{(q_\varkappa \xi_c)^{2(d-3)}}
\left(\frac{\xi_0}{\xi_c}\right)^{d-2-z}
\left(\frac{E-E_c}{g_*^{\rm n} E_c}\right)^{1+\Delta_2^{\rm n}/d},
\label{eq39b}
\end{equation}
such that
$z^{\rm n}_\phi = {d^2}/({d+\Delta_2^{\rm n}})$.

In particular, for $d=2+\epsilon$ we find
$z^{\rm n}_\phi \simeq 2(1+\sqrt{\epsilon/2})$. The contributions of domains (I) and (II) to the dephasing rate scale
as $[\xi(E)]^{d-4}\sim (E-E_c)^{(4-d) \nu_{\rm n}}$ for the case (i) and $[\xi(E)]^{-d-\Delta_2^{\rm n}}\sim (E-E_c)^{(d+\Delta_2^{\rm n}) \nu_{\rm n}}$ for the case (ii). They are therefore subleading for $\nu_{\rm n}>1/d$ in comparison with the results \eqref{eq39a} and \eqref{eq39b}, respectively.

The third contribution in Eq.~(\ref{eq14b}) can be important only in $d=2+\epsilon$ when $\xi_c\ll\xi_0$. Since in $d=2+\epsilon$ the following inequality holds: $d+1+\Delta_2+\Delta_2^{\rm n}>0$, the integral over momenta and frequencies is dominated by $q\sim \xi_0^{-1}$ and $\omega\sim E-E_c$, respectively (region (III)). Then, we find,
\begin{gather}
\xi_0^{d-2} \Delta
\int\limits_{\omega_\xi}^{E-E_c}d\omega \int\limits_{1/\xi_0}^{1/\xi_c} dq\
q^{d-3} (q\xi_0)^{-d+2-\Delta_2}\re \mathcal{D}_c(\omega,q) \notag \\
\sim
E_c \left (\frac{\xi_0}{\xi_c}\right )^{d+\Delta_2^{\rm n}} \left (\frac{E-E_c}{g_*^{\rm n} E_c} \right )^{1+\Delta_2^{\rm n}/d} .
\label{eqAdd}
\end{gather}
This contribution has the same power law dependence on $E-E_c$ as the result \eqref{eq39b}. Since $q_\varkappa\ll1/\xi_0\ll1/\xi_c$ and $z+2(d-2)-\Delta_2^{\rm n}>0$, we find $(\xi_0/\xi_c)^{d+\Delta_2^{\rm n}} \ll
(q_\varkappa \xi_c)^{2(d-3)}(\xi_0/\xi_c)^{d-2-z}$. Therefore, the contribution \eqref{eqAdd} is always smaller than
the result \eqref{eq39b} and the region $1/\xi_0<q<1/\xi_c$ does not influence the results \eqref{eq39a} and \eqref{eq39b} for the dephasing rate.

In $d=3$ the momentum scale $q_\varkappa=0$ and $\xi_c\sim\xi_0$. The dominant contribution to the dephasing rate comes from momenta $q\sim q_\omega$. Therefore, the dephasing rate is given by Eq. \eqref{eq27} with $d=3$ such that $z^{\rm n}_\phi = 9$ (for the case $\alpha=1$).
We summarize the results for the critical exponent $z^{\rm n}_\phi$ in Table \ref{Tab_zP}.

Finally, it is worth noticing that in the model without Coulomb interaction
between localized particles, where the conventional Mott formula applies,
the dephasing rate contains an extra power of $(E-E_c)/E_c$, as compared to Eqs. (\ref{eq27}) and (\ref{eq39a}). This leads to the following results: $z^{\rm n}_\phi=d^2/4$ for the case (i) and $z^{\rm n}_\phi=d^2/(2d+\Delta_2^{\rm n})$ for the case (ii).
We summarize the results for the exponent $z^{\rm n}_\phi$ in Table \ref{Tab_zP}.


\begin{table}[b]
\caption{Results for the exponent $z^{\rm n}_\phi$}
\label{Tab_zP}
\begin{tabular}{l|cccc}
& \hspace{0.5cm} &$\alpha=1$ & \hspace{0.5cm} & $\alpha=2$ \\
\hline
$4-2d-\Delta_2^{\rm n} <0$ &\hspace{0.5cm} &  $\displaystyle {d^2}/({4-d})$ & \hspace{0.5cm} &$\displaystyle {d^2}/{4}$\\
$4-2d-\Delta_2^{\rm n} \geqslant 0$ & \hspace{0.5cm} &$\displaystyle {d^2}/({d+\Delta_2^{\rm n}})$ & \hspace{0.5cm} &$\displaystyle {d^2}/({2d+\Delta_2^{\rm n}})$
\end{tabular}
\end{table}


\end{document}